\documentclass[journal]{IEEEtran}
\IEEEoverridecommandlockouts
\usepackage{amsmath,amssymb,amsfonts}
\usepackage[linesnumbered,ruled,vlined]{algorithm2e}
\usepackage{graphicx}
\usepackage[noadjust]{cite}
\usepackage{textcomp}
\usepackage{xcolor}
\usepackage{newtxmath}
\usepackage{nccmath}

\usepackage{amsmath}
\usepackage{tikz}
\usepackage{booktabs}
\usepackage{array}
\usepackage{multirow}
\usepackage{enumitem}
\usepackage{url}

\makeatletter\def\@IEEEfigurecaptionsepspace{\vskip 3pt\relax}\makeatother

\usepackage{amsthm}

\newtheorem{theorem}{Theorem}
\newtheorem{lemma}{Lemma}

\newtheorem{proposition}{Proposition}

\newtheorem{remark}{Remark}

\newtheorem{assumption}{Assumption}
\begin{document}
\title{Bisect-Index-Choose: A Fast and Exact Algorithm for EV Charging under Minimum Charging Rates}
\author{
Can~Berk~Saner,~\IEEEmembership{Member,~IEEE}
        \thanks{C.B. Saner is with the Department of Engineering, Newcastle University, Newcastle upon Tyne, NE1 7RU, U.K. (e-mail: ncbs4@newcastle.ac.uk).}
        }

\maketitle

\begin{abstract}
Electric vehicle (EV) charging stations are increasingly operated with an aggregate power limit below the sum of their port ratings, and require scheduling algorithms to divide the available power among the connected EVs. Charging standards permit a current that is either zero or at least a minimum rate, typically 6~A, which makes this allocation problem non-convex. Existing methods either solve a mixed-integer program with a commercial solver, which can be computationally expensive, adds software dependencies, and incurs licensing costs, or apply heuristics without optimality guarantees. This paper proposes Bisect-Index-Choose (BIC), a fast and solver-free scheduling algorithm that computes a globally optimal allocation among EVs under mild assumptions on their utilities. For $n$ EVs, BIC runs in $O(n \log n)$ time when all EVs share one maximum rate and in $O(n^2)$ time for two, times a logarithmic factor in the numerical precision. In tests with up to 1000 EVs, BIC attains global optimality and is up to three orders of magnitude faster than the solver. On real-world charging data, it needs less capacity and delivers more energy than existing heuristics. We implement BIC on a low-cost microcontroller and show that it solves instances with 1000 EVs in less than 90~ms.
\end{abstract}


\section{Introduction}\label{sec:intro}

Electric vehicle (EV) sales grew by a fifth in 2025 to surpass 20 million, and the electricity drawn for charging is projected to grow around sixfold by 2035~\cite{gevo2026}. Amid this growth, charging stations at workplaces, depots, and public lots, which host many charging ports, are increasingly operated in an oversubscribed configuration~\cite{lee2021adaptive, cho2026distributed}, where the station's aggregate power limit falls below the sum of its ports' rated powers. This cap may come from a physical limit, such as that of the feeding transformer or cable, or from an exogenous setpoint imposed by a higher-level energy management system or aggregator~\cite{wang2025online}. Left uncoordinated, the combined charging demand may exceed this cap and thereby overload the electrical infrastructure or violate the prescribed operating limit. To set the charging powers of the connected EVs under this capacity constraint, a wide variety of EV charging scheduling algorithms has been proposed~\cite{nimalsiri2020survey, dahiwale2024comprehensive}.

A common assumption in this literature is that the charging rate of each EV, expressed as either current or power, can vary continuously between zero and its maximum~\cite{nimalsiri2020survey, liang2025online, rudnik2020real}. Such an assumption is convenient because the resulting allocation can be cast as a convex optimization problem and solved efficiently to global optimality (e.g.,~\cite{lee2021adaptive, chen2021smoothed}). In practice, however, EV charging hardware does not support this continuous range~\cite{digiorgio2014electric, ireshika2022iec, nomura2026electric}. In particular, EV charger standards such as IEC~61851~\cite{iec61851} and SAE~J1772~\cite{saej1772} specify that the control pilot allows the charging current to be either zero or within a valid range bounded below by a minimum threshold, typically 6~A, and above by the maximum charger rating. Consequently, allocations computed under the continuous-rate model may assign charging rates that the EV cannot execute, which makes them infeasible in practice~\cite{qian2023impact}.

Introducing the minimum charging rate renders the allocation problem non-convex, as the admissible charging rates of each EV now form a disconnected set, and standard convex optimization methods no longer apply directly. Methods that account for minimum charging rates fall into two approaches.

The first approach introduces binary variables to represent the charging status of each EV, which leads to mixed-integer programming (MIP) models~\cite{digiorgio2014electric, ireshika2022iec, qian2023impact, rudnik2020real, frendo2019real, cavalcantilauro2026scheduling}. Exact MIP solution methods certify global optimality, but their worst-case computation time grows exponentially with the number of binary variables~\cite{rudnik2020real}, and solution times that exceed real-time requirements have been reported for a few tens of EVs~\cite{frendo2019real, hegele2023efficient}. They also rely on specialized optimization software, which entails memory, runtime, and dependency overheads that complicate deployment on embedded controllers~\cite{hegele2023efficient, uiterkamp2025scheduling}, and commercial solvers add licensing costs.

The second approach avoids the MIP and uses tractable heuristics and approximation methods. These include rounding of continuous solutions~\cite{lee2021adaptive, kisacikoglu2018distributed}, priority rules for on/off charging that serve the most urgent EVs first~\cite{zeballos2019proportional, xu2016dynamic}, greedy allocation~\cite{hegele2023efficient}, and approximation algorithms with performance bounds~\cite{khonji2018approximation}. Such methods are cheaper to compute, and some run on embedded controllers~\cite{kisacikoglu2018distributed, hegele2023efficient}, but do not generally guarantee global optimality.

Exact methods that avoid a general-purpose solver have been developed for special cases of the problem. When the admissible rates are discrete, evenly spaced and include zero, a greedy method finds the optimal allocation, also under phase unbalance constraints~\cite{saner2025optimal}. The minimum rate is handled exactly for a single EV over its parking time~\cite{uiterkamp2018offline}, and more generally for resource allocation problems with disjoint interval bound constraints whose objective is a sum of shifts $\phi(x_i + b_i)$ of one convex function $\phi$~\cite{uiterkamp2025symmetric}. Neither method applies when the utilities differ by a scaling, for example in the smoothed least-laxity-first (sLLF) algorithm~\cite{chen2021smoothed}, which scales the utility of each EV by its own maximum rate.

To provide a fast exact solution for this setting, this paper proposes Bisect-Index-Choose (BIC), a solver-free algorithm that computes a globally optimal allocation among the EVs at a charging station. The method applies when the charging scheduler maximizes the sum of EV utilities in each allocation step, under mild assumptions on the utilities that are satisfied by several common utility families. The main contributions of this work are summarized as follows:

\begin{itemize}
\item \emph{Global optimality.} We prove that the BIC algorithm returns a globally optimal allocation whenever the utilities are continuously differentiable, strictly increasing and strictly concave, and the EVs with the same maximum rate exhibit a priority order (Assumption~\ref{as:priority}).
\item \emph{Low computational complexity.} For $n$ connected EVs, BIC runs in $O(n \log n)$ time when all EVs share one maximum rate and in $O(n^2)$ time for two maximum rates, times a logarithmic factor in the numerical precision.
\item \emph{Solver-free implementation.} The algorithm uses only elementary operations, such as sorting, scalar bisection, and arithmetic, and does not require a mixed-integer optimization solver, which simplifies deployment on resource-constrained devices and avoids solver licensing costs.
\item \emph{Empirical validation.} We validate the optimality and computation time of BIC against a commercial MIP solver on instances with up to 1000 EVs, and compare its solution quality with two heuristics adapted from the literature, both on synthetic instances and on real-world charging sessions. Furthermore, we implement BIC on a low-cost microcontroller and show that it solves instances with up to 1000 EVs in a fraction of a control period.\footnote{\label{fn:code}The implementation of BIC, the code to reproduce all results, and a formalization of theorems and lemmas in the Lean 4 proof assistant are available at \url{https://github.com/canberksaner/bic-public}.}
\end{itemize}

The remainder of the paper is organized as follows. Section~\ref{sec:formulation} introduces the system model and formulates the per-period allocation problem. Section~\ref{sec:relaxed} reviews its continuous relaxation. Section~\ref{sec:bic} develops BIC and establishes its optimality and computational complexity. Section~\ref{sec:setup} describes the experimental setup, Section~\ref{sec:numerics} presents and discusses the results, and Section~\ref{sec:conclusion} concludes the paper.

\section{System Model and Problem Formulation}\label{sec:formulation}

We consider an EV charging station with multiple charging ports. EVs arrive over time, occupy a port, and are charged until they depart. To control the charging process, we adopt a discrete-time model: time is divided into control periods of equal length, indexed by $t$, and the charging rates are held constant within each period. The period length is a design choice, from a few minutes to an hour~\cite{lee2021adaptive, li2025out, qian2023impact, liang2025online, wang2025online}. We write $\mathcal{E}_t$ for the set of EVs connected at period $t$, and $x_{i,t}$ for the charging rate of EV $i$, expressed as power in kW, which the station decides at each period. Each rate must lie in the set $\mathcal{R}_i$ of admissible charging rates of EV $i$, specified below. The total power drawn by the station is capped by the power budget $B_t$, which may change from period to period, for example during a demand response event.

This decision must be made with limited information. When an EV arrives, the station learns its parameter vector $\boldsymbol{\theta}_{i,t}$, which may contain declared or measured quantities, such as the departure time and the remaining energy demand, as well as derived ones, such as urgency or laxity~\cite{chen2021smoothed}. These entries may be updated as time progresses. Future arrivals, however, are unknown~\cite{li2025out}, and so are the power budgets $B_t$ of future periods. The station therefore operates online: at each control period, a central scheduler takes what is currently visible, namely the connected EVs, their parameters, and the power budget, and returns feasible charging rates, which the station applies until the next period. We denote this scheduler by \textsc{Scheduler} and refer to this loop as the \textit{online operation} of the station.

Since each control period is treated identically, the rest of the paper concerns a single call of \textsc{Scheduler}: we fix one period and drop the subscript $t$ hereafter. We write $\mathcal{N} = \{1, \dots, n\}$ for the connected EVs, $B > 0$ for the power budget, $\boldsymbol{\theta}_i$ for the parameters, and $x_i$ for the charging rates.

As discussed in Section~\ref{sec:intro}, the charging standards restrict the admissible set of EV $i$ to
\begin{equation}\label{eq:R}
\mathcal{R}_i = \{0\} \cup [x^{\min}, x_i^{\max}], \qquad 0 < x^{\min} < x_i^{\max},
\end{equation}
where $x^{\min}$ is the standardized minimum charging rate and $x_i^{\max}$ is the maximum rate of EV $i$, given by the smaller of the port rating and the rating of the vehicle's onboard charger. The minimum rate is common to all EVs, because the standards fix the minimum current and we assume that all EVs are supplied in the same configuration (e.g.\ three-phase at 0.4~kV).

The maximum rates of the connected EVs take $K$ distinct values $\bar{x}_1 < \dots < \bar{x}_K$. We group the EVs accordingly. The \emph{group} $\mathcal{N}_k$ is the set of $n_k$ EVs with maximum rate $\bar{x}_k$, so that $x_i^{\max} = \bar{x}_k$ for all $i \in \mathcal{N}_k$, and we also refer to it as group~$k$. In practice, $K$ is small, which keeps the search of BIC short (Section~\ref{sec:algorithm}). Over 90\% of the passenger cars and SUVs in the OpenEV database~\cite{openev} (release v1.24.0, current production, CCS2 inlet, battery of at least 30~kWh) have an AC charging rating of 11 or 22~kW.

We instantiate \textsc{Scheduler} by utility maximization, a common approach in network resource allocation and smart charging~\cite{kelly1998rate, samadi2010optimal, theodoropoulos2022proportionally, chen2021smoothed}. Each connected EV $i \in \mathcal{N}$ is assigned a utility function $f_i$, constructed from its parameter vector $\boldsymbol{\theta}_i$, such that $f_i(x_i)$ quantifies the benefit of charging EV $i$ at rate $x_i$ during the current period. \textsc{Scheduler} then allocates the power budget to maximize the total utility of the EVs:
\begin{equation}\label{eq:P}
\begin{aligned}
\max_{\boldsymbol{x} \in \mathbb{R}^n} \quad & \sum_{i \in \mathcal{N}} f_i(x_i) \\
\text{s.t.} \quad & \sum_{i \in \mathcal{N}} x_i \le B, \ x_i \in \mathcal{R}_i, \ \forall i \in \mathcal{N}.
\end{aligned}
\tag{$P$}
\end{equation}
The disconnected set~\eqref{eq:R} makes \eqref{eq:P} non-convex. We assume $f_i(0) = 0$ for all $i \in \mathcal{N}$, so that an EV with $x_i = 0$ contributes nothing to the objective. This is without loss of generality, since adding a constant to $f_i$ does not change the maximizers of~\eqref{eq:P}, and the assumptions below refer to these normalized utilities. We place two assumptions on the utilities.

\begin{assumption}[Regularity]\label{as:regularity}
Each $f_i$ is continuously differentiable, strictly increasing, and strictly concave on $[0, x_i^{\max}]$.
\end{assumption}

\begin{assumption}[Priority order]\label{as:priority}
The EVs of each group $\mathcal{N}_k$ can be labeled so that
\begin{equation}\label{eq:order}
i \le j,\ i, j \in \mathcal{N}_k \ \Longrightarrow \ f_i(x) \ge f_j(x) \quad \forall x \in [x^{\min}, \bar{x}_k].
\end{equation}
In words, no two utility curves of the same group cross on its operating range. No order is required across groups.
\end{assumption}

Appendix~\ref{app:assumptions} gives a general condition under which a family of utilities satisfies both assumptions, and Table~\ref{tab:families} lists common examples. In each example, the utilities of a group differ only in a scalar key, such as a weight or the laxity, and the priority order sorts the EVs by this key.

Since each $f_i$ is increasing, the allocation $x_i = x_i^{\max}$ for all $i \in \mathcal{N}$ is optimal if $B \ge \sum_{i \in \mathcal{N}} x_i^{\max}$. We therefore assume $B < \sum_{i \in \mathcal{N}} x_i^{\max}$ hereafter.

\section{Solving the Relaxed Problem}\label{sec:relaxed}

Before turning to \eqref{eq:P}, we briefly review the solution of the relaxed problem, in which the minimum rate is ignored:
\begin{equation}\label{eq:P_bar}
\begin{aligned}
\max_{\boldsymbol{x} \in \mathbb{R}^n} \quad & \sum_{i \in \mathcal{N}} f_i(x_i) \\
\text{s.t.} \quad & \sum_{i \in \mathcal{N}} x_i \le B, \ x_i \in [0, x_i^{\max}], \ \forall i \in \mathcal{N}.
\end{aligned}
\tag{$\bar{P}$}
\end{equation}
Problem \eqref{eq:P_bar} is the continuous-rate model common in the literature, including the sLLF algorithm of~\cite{chen2021smoothed}, and it is also the basis of our algorithm.

Problem \eqref{eq:P_bar} is convex, and the Karush-Kuhn-Tucker (KKT) conditions are necessary and sufficient for optimality. We dualize only the budget constraint, with a multiplier $\lambda \ge 0$. The Lagrangian then separates across the EVs, and for a fixed $\lambda$, each EV $i \in \mathcal{N}$ solves
\begin{equation}\label{eq:pointwise}
x_i(\lambda) := \arg\max_{x_i \in [0, x_i^{\max}]} \ \bigl\{f_i(x_i) - \lambda x_i\bigr\}.
\end{equation}
The unconstrained maximizer satisfies $f_i'(x_i) = \lambda$. Its projection onto $[0, x_i^{\max}]$ is
\begin{equation}\label{eq:selector}
x_i(\lambda) = \Bigl[(f_i')^{-1}(\lambda)\Bigr]_0^{x_i^{\max}},
\end{equation}
where $[z]_a^b := \min\{\max\{z, a\}, b\}$. The inverse $(f_i')^{-1}$ is well defined on $[f_i'(x_i^{\max}), f_i'(0)]$ by Assumption~\ref{as:regularity}.

Write $\boldsymbol{x}(\lambda) := (x_i(\lambda))_{i \in \mathcal{N}}$ for the allocation under multiplier $\lambda$, and define its aggregate as $S(\lambda) := \sum_{i \in \mathcal{N}} x_i(\lambda)$. Since each $x_i(\lambda)$ is continuous and non-increasing in $\lambda$, the aggregate $S(\lambda)$ is also continuous and non-increasing. In addition, $S(0) = \sum_{i \in \mathcal{N}} x_i^{\max} > B$ and $S(\lambda) = 0$ for $\lambda \ge \max_{i \in \mathcal{N}} f_i'(0)$. Therefore, there exists $\lambda^\star \ge 0$ such that $S(\lambda^\star) = B$, and this value can be computed efficiently by bisection on $\lambda$ over the bracket $[0, \max_{i \in \mathcal{N}} f_i'(0)]$. The corresponding allocation $\boldsymbol{x}(\lambda^\star)$ is the unique maximizer of \eqref{eq:P_bar}, by strict concavity.

\section{The Bisect-Index-Choose (BIC) Algorithm for Solving \eqref{eq:P}}\label{sec:bic}

We now return to \eqref{eq:P}. Its admissible set $\mathcal{R}_i$ is not convex, so the KKT conditions cannot be used directly. We instead follow the generalized Lagrange multiplier method of~\cite{everett1963generalized}: we dualize the budget constraint with a multiplier $\lambda \ge 0$ as before, but keep the non-convex set $\mathcal{R}_i$, and redefine $x_i(\lambda)$ and $S(\lambda)$ for~\eqref{eq:P}: for each $i \in \mathcal{N}$,
\begin{equation}\label{eq:chi_argmax}
x_i(\lambda) \in \arg\max_{x_i \in \mathcal{R}_i} \ \bigl\{f_i(x_i) - \lambda x_i\bigr\}.
\end{equation}
As the feasible set is the union of two disjoint sets, the maximization reduces to a comparison between two candidates: the point $0$, with objective value $0$, and the best point of the interval $[x^{\min}, x_i^{\max}]$, namely the projection $\bigl[(f_i')^{-1}(\lambda)\bigr]_{x^{\min}}^{x_i^{\max}}$. With the threshold
\begin{equation}\label{eq:lambda_th}
\lambda^{\mathrm{th}}_i := \frac{f_i(x^{\min})}{x^{\min}},
\end{equation}
the comparison gives (see Appendix~\ref{app:closed} for the derivation)
\begin{equation}\label{eq:chi}
x_i(\lambda) =
\begin{cases}
\bigl[(f_i')^{-1}(\lambda)\bigr]_{x^{\min}}^{x_i^{\max}}, & \lambda < \lambda^{\mathrm{th}}_i, \\
0, & \lambda \ge \lambda^{\mathrm{th}}_i.
\end{cases}
\end{equation}
Each $x_i(\lambda)$ is non-increasing in $\lambda$ and drops from at least $x^{\min}$ to $0$ at $\lambda = \lambda^{\mathrm{th}}_i$. Hence, the aggregate $S(\lambda) = \sum_{i \in \mathcal{N}} x_i(\lambda)$ is non-increasing in $\lambda$, but no longer continuous. Its jumps create gaps in its range, which we refer to as \emph{Everett gaps} after~\cite{everett1963generalized}.

\begin{table}[t]
\centering
\caption{Utility functions that satisfy Assumptions~\ref{as:regularity} and~\ref{as:priority}, for an EV $i$ of group $\mathcal{N}_k$.}
\label{tab:families}
\footnotesize
\setlength{\tabcolsep}{2.5pt}
\begin{tabular}{llll}
\toprule
Utility & $f_i(x)$ & Conditions & Priority order \\
\midrule
sLLF~\cite{chen2021smoothed} & $\bar{x}_k \log(\ell_i - 1 + x/\bar{x}_k)$\,$^\dag$ & $\ell_i > 1$ & ascending $\ell_i$ \\
Quadratic~\cite{samadi2010optimal} & $a_i x - \frac{b}{2}x^2$ & $b > 0$, $a_i > b\,\bar{x}_k$ & descending $a_i$ \\
Logarithmic & $a_i \log\bigl(1 + \frac{x}{b}\bigr)$ & $a_i, b > 0$ & descending $a_i$ \\
Exponential & $a_i\bigl(1 - e^{-b x}\bigr)$ & $a_i, b > 0$ & descending $a_i$ \\
\bottomrule
\multicolumn{4}{p{0.94\columnwidth}}{\scriptsize $^\dag$ $\ell_i$ is the laxity of EV $i$ in control periods (Section~\ref{sec:setup-utility}). Displayed up to the additive constant $-\bar{x}_k \log(\ell_i - 1)$, which enforces $f_i(0) = 0$.}
\end{tabular}
\end{table}

\subsection{The Bisection Phase and Fast-Path}\label{sec:bisection}

Although $S$ may be discontinuous, it remains non-increasing, so bisection on $\lambda$ over $[0, L]$, where $L := \max_{i \in \mathcal{N}} \lambda^{\mathrm{th}}_i$, still finds a multiplier $\lambda^\star$ with $S(\lambda^\star) = B$ whenever one exists. The corresponding allocation is globally optimal, despite the non-convexity. This is stated below.

\begin{theorem}\label{thm:bisection}
Assume there exists $\lambda^\star \ge 0$ such that $S(\lambda^\star) = B$. Then $\boldsymbol{x}(\lambda^\star)$ is a global maximizer of \eqref{eq:P}.
\end{theorem}

\begin{proof}
Let $\tilde{\boldsymbol{x}}$ be any feasible solution to \eqref{eq:P} with $\sum_i \tilde{x}_i = \tilde{B} \le B$. By the definition of $x_i(\lambda^\star)$ as a maximizer in \eqref{eq:chi_argmax}, $f_i(x_i(\lambda^\star)) - \lambda^\star x_i(\lambda^\star) \ge f_i(\tilde{x}_i) - \lambda^\star \tilde{x}_i$ for all $i \in \mathcal{N}$. Summing over $i$ and using $S(\lambda^\star) = B$ gives
\begin{equation*}
\sum_{i \in \mathcal{N}} f_i(x_i(\lambda^\star)) - \lambda^\star B \ \ge\ \sum_{i \in \mathcal{N}} f_i(\tilde{x}_i) - \lambda^\star \tilde{B}.
\end{equation*}
Hence $\sum_i f_i(x_i(\lambda^\star)) \ge \sum_i f_i(\tilde{x}_i) - \lambda^\star(\tilde{B} - B) \ge \sum_i f_i(\tilde{x}_i)$, since $\tilde{B} \le B$ and $\lambda^\star \ge 0$. Therefore $\boldsymbol{x}(\lambda^\star)$ solves \eqref{eq:P}.
\end{proof}

If the bisection phase returns such a value $\lambda^\star$, the algorithm terminates with the globally optimal solution $\boldsymbol{x}(\lambda^\star)$. We call this early exit the \emph{fast-path}. Otherwise, no multiplier attains the budget: the aggregate jumps from above $B$ to below $B$, i.e., $B$ lies inside an Everett gap, and the algorithm proceeds to the subsequent phases.

\subsection{The Indexing Phase}\label{sec:indexing}

In the case where no value of $\lambda$ satisfies $S(\lambda) = B$, no single multiplier determines the optimal allocation, so we determine it through the set of EVs that are charged. We call the set of EVs that are switched on, i.e., charged at a positive rate, the \emph{on-set} of a feasible allocation. By~\eqref{eq:R}, every EV in the on-set charges at a rate of at least $x^{\min}$. Once the on-set is fixed, each EV in it charges within its interval $[x^{\min}, x_i^{\max}]$, every other EV is at zero, and the remaining problem is convex. To formalize this, we define the following optimization problem, in which the on-set is fixed to $\mathcal{I} \subseteq \mathcal{N}$:
\begin{equation}\label{eq:P_J}
\begin{aligned}
\max_{\boldsymbol{x} \in \mathbb{R}^{|\mathcal{I}|}} \quad & \sum_{i \in \mathcal{I}} f_i(x_i) \\
\text{s.t.} \quad & \sum_{i \in \mathcal{I}} x_i \le B, \ x_i \in [x^{\min}, x_i^{\max}], \ \forall i \in \mathcal{I}.
\end{aligned}
\tag{$P_{\mathcal{I}}$}
\end{equation}
We call \eqref{eq:P_J} the \emph{restriction} of \eqref{eq:P} to $\mathcal{I}$. For each fixed $\mathcal{I}$, it is a convex program, since its feasible set is convex and its objective is separable and concave. Hence \eqref{eq:P_J} can be solved by bisection on the multiplier, exactly as the relaxed problem, with the projection now taken onto the interval $[x^{\min}, x_i^{\max}]$. We denote by $V(\mathcal{I})$ the optimal value of \eqref{eq:P_J} and set $V(\emptyset) := 0$. Since every EV in $\mathcal{I}$ charges at least $x^{\min}$, \eqref{eq:P_J} is feasible if and only if $|\mathcal{I}|\,x^{\min} \le B$.

Solving \eqref{eq:P} therefore amounts to finding a feasible on-set with the largest value, $\mathcal{I}^\star \in \arg\max \{V(\mathcal{I}) : \mathcal{I} \subseteq \mathcal{N},\ |\mathcal{I}|\,x^{\min} \le B\}$, and solving $(P_{\mathcal{I}^\star})$. A brute-force enumeration of all $2^n$ subsets is, however, intractable. We therefore search the on-sets systematically, by the number of EVs they switch on in each group. For an on-set $\mathcal{I}$, let $c_k$ be the number of its EVs in group $k$, i.e., $c_k = |\mathcal{I} \cap \mathcal{N}_k|$, and call $\boldsymbol{c} = (c_1, \dots, c_K)$ its \emph{cardinality vector}. For example, an on-set that switches on five 11~kW EVs and three 22~kW EVs has $\boldsymbol{c} = (5, 3)$. Two constraints bound $\boldsymbol{c}$: group $k$ contains only $n_k$ EVs, and every switched-on EV needs at least $x^{\min}$ of the budget, so at most $n^{\max} := \lfloor B / x^{\min} \rfloor$ EVs can be switched on. The admissible cardinality vectors are therefore
\begin{equation}\label{eq:cardinalities}
\mathcal{C} := \Bigl\{\boldsymbol{c} \in \mathbb{Z}^K : 0 \le c_k \le n_k \ \forall k, \ \sum_{k} c_k \le n^{\max}\Bigr\}.
\end{equation}
For $\boldsymbol{c} \in \mathcal{C}$, let $\mathcal{F}_{\boldsymbol{c}} := \{\mathcal{I} \subseteq \mathcal{N} : |\mathcal{I} \cap \mathcal{N}_k| = c_k \ \forall k\}$ be the family of on-sets with cardinality vector $\boldsymbol{c}$, and let $V_{\boldsymbol{c}} := \max_{\mathcal{I} \in \mathcal{F}_{\boldsymbol{c}}} V(\mathcal{I})$ be the best value in this family. Every feasible on-set belongs to exactly one family $\mathcal{F}_{\boldsymbol{c}}$ with $\boldsymbol{c} \in \mathcal{C}$, hence
\begin{equation}\label{eq:decomposition}
V(\mathcal{I}^\star) \;=\; \max_{\boldsymbol{c} \in \mathcal{C}} \, \max_{\mathcal{I} \in \mathcal{F}_{\boldsymbol{c}}} V(\mathcal{I}) \;=\; \max_{\boldsymbol{c} \in \mathcal{C}} V_{\boldsymbol{c}}.
\end{equation}
We therefore find $\mathcal{I}^\star$ in two steps:
\begin{enumerate}
\item[(i)] for a given $\boldsymbol{c} \in \mathcal{C}$, find a best on-set in $\mathcal{F}_{\boldsymbol{c}}$, i.e., one that attains $V_{\boldsymbol{c}}$;
\item[(ii)] find the best cardinality vector, $\boldsymbol{c}^\star \in \arg\max_{\boldsymbol{c} \in \mathcal{C}} V_{\boldsymbol{c}}$, and take $\mathcal{I}^\star$ to be a best on-set in $\mathcal{F}_{\boldsymbol{c}^\star}$, so that $V(\mathcal{I}^\star) = V_{\boldsymbol{c}^\star}$.
\end{enumerate}
Theorem~\ref{thm:prefix} shows that step~(i) requires no search: the set $\mathcal{I}^\star_{\boldsymbol{c}}$ of the first $c_k$ EVs of each group $\mathcal{N}_k$ in the priority order~\eqref{eq:order} is a best on-set in $\mathcal{F}_{\boldsymbol{c}}$, so $V_{\boldsymbol{c}} = V(\mathcal{I}^\star_{\boldsymbol{c}})$.

\begin{theorem}\label{thm:prefix}
Let $\boldsymbol{c} \in \mathcal{C}$, and let $\mathcal{I}^\star_{\boldsymbol{c}}$ consist of the first $c_k$ EVs of each group $\mathcal{N}_k$ in the priority order~\eqref{eq:order}. Then
\begin{equation*}
V(\mathcal{I}^\star_{\boldsymbol{c}}) \ \ge\ V(\mathcal{I}) \qquad \forall \mathcal{I} \in \mathcal{F}_{\boldsymbol{c}}.
\end{equation*}
In words, among all on-sets that switch on the same number of EVs in each group, the on-set that switches on the highest-priority EVs of each group yields the largest utility.
\end{theorem}

\begin{proof}
See Appendix~\ref{app:thm2}.
\end{proof}

The \emph{indexing phase} sorts the EVs of each group $\mathcal{N}_k$ in the priority order~\eqref{eq:order}. For the families of Table~\ref{tab:families}, sorting the EVs amounts to sorting their keys within each group. After the sort, step~(i) reduces to reading off the first $c_k$ EVs of each group, and step~(ii) is the task of the \emph{choosing phase}.

\subsection{The Choosing Phase}\label{sec:choosing}

\begin{algorithm}[t]
\DontPrintSemicolon
\KwIn{groups $\mathcal{N}_1, \mathcal{N}_2$ sorted in the priority order, with $n_1$ and $n_2$ EVs; the largest number $n^{\max}$ of EVs that can be switched on}
\KwOut{best cardinality vector $\boldsymbol{c}^\star$}
$V_{\mathrm{best}} \leftarrow -\infty$\;
$c_2 \leftarrow n_2$\nllabel{ln:ch-init}\;
\For{$c_1 \leftarrow 0$ \KwTo $\min(n_1, n^{\max})$\nllabel{ln:ch-for}}{
    $c_2 \leftarrow \min(c_2,\ n^{\max} - c_1)$\nllabel{ln:ch-cap}\;
    \While{$c_2 > 0$\nllabel{ln:ch-while}}{
        \lIf{$V_{(c_1, c_2-1)} < V_{(c_1, c_2)}$\nllabel{ln:ch-test}}{\textbf{break}}
        $c_2 \leftarrow c_2 - 1$\nllabel{ln:ch-dec}\;
    }
    $\hat c_2(c_1) \leftarrow c_2$\nllabel{ln:ch-rec}\;
    \If{$V_{(c_1, \hat c_2(c_1))} > V_{\mathrm{best}}$\nllabel{ln:ch-if}}{
        $V_{\mathrm{best}} \leftarrow V_{(c_1, \hat c_2(c_1))}$;\quad $\boldsymbol{c}_{\mathrm{best}} \leftarrow (c_1, \hat c_2(c_1))$\nllabel{ln:ch-keep}\;
    }
}
$\boldsymbol{c}^\star \leftarrow \boldsymbol{c}_{\mathrm{best}}$\;
\Return{$\boldsymbol{c}^\star$}\;
\caption{Choosing phase for two groups, $\textsc{Choose}(\mathcal{N}_1, \mathcal{N}_2, n^{\max})$.}
\label{alg:choose}
\end{algorithm}

The choosing phase finds $\boldsymbol{c}^\star$. Evaluating $V_{\boldsymbol{c}}$ for every $\boldsymbol{c} \in \mathcal{C}$ would need up to $\prod_k (n_k + 1)$ restrictions. As we show next, far fewer are enough. We write $\boldsymbol{e}_k$ for the $k$-th unit vector, and $\boldsymbol{c}(k; q)$ for the vector $\boldsymbol{c}$ with its $k$-th entry replaced by $q$. For $\boldsymbol{c} \in \mathcal{C}$ and a group $k$, let
\begin{equation*}
\hat c_k(\boldsymbol{c}) := \min \arg\max_{q} \bigl\{ V_{\boldsymbol{c}(k; q)} : \boldsymbol{c}(k; q) \in \mathcal{C} \bigr\}
\end{equation*}
be the best count of group $k$. Intuitively, $\hat c_k(\boldsymbol{c})$ answers the following question: \emph{if the numbers of EVs switched on in all other groups stay as in $\boldsymbol{c}$, what is the smallest number of EVs of group $k$ that gives the largest value?} For $K = 2$, $\hat c_2(\boldsymbol{c})$ depends only on $c_1$, and we write $\hat c_2(c_1)$. The following theorem gives two properties of the best count $\hat c_k(\boldsymbol{c})$, which allow an efficient search for $\boldsymbol{c}^\star$.

\begin{theorem}\label{thm:dr}
Let $\boldsymbol{c} \in \mathcal{C}$, and let $k$ be a group.
\begin{enumerate}
\item[(a)] If $c_k \ge 1$, then $V_{\boldsymbol{c}} > V_{\boldsymbol{c} - \boldsymbol{e}_k} \iff \hat c_k(\boldsymbol{c}) \ge c_k$.
\item[(b)] If $l \ne k$ is another group and $\boldsymbol{c} + \boldsymbol{e}_l \in \mathcal{C}$, then $\hat c_k(\boldsymbol{c} + \boldsymbol{e}_l) \le \hat c_k(\boldsymbol{c})$.
\end{enumerate}
In words, switching on the $c_k$-th EV of a group raises the value exactly when the best count of that group is at least $c_k$, and this best count does not rise when one more EV of another group is switched on.
\end{theorem}

\begin{proof}
See Appendix~\ref{app:dr}.
\end{proof}

Theorem~\ref{thm:dr} gives rise to the choosing-phase algorithm \textsc{Choose}, stated for $K = 2$ in Algorithm~\ref{alg:choose}. The algorithm works as follows. The aim is to find $\boldsymbol{c}^\star \in \arg\max_{\boldsymbol{c} \in \mathcal{C}} V_{\boldsymbol{c}}$, which is equivalent to finding $\arg\max_{c_1} V_{(c_1, \hat c_2(c_1))}$. The outer loop (line~\ref{ln:ch-for}) scans $c_1$ upwards, from $0$ to its highest feasible value. For each $c_1$, it finds $\hat c_2(c_1)$ and records $V_{(c_1, \hat c_2(c_1))}$.

\begin{algorithm}[t]
\DontPrintSemicolon
\KwIn{budget $B$, utilities $\{f_i\}_{i \in \mathcal{N}}$, minimum rate $x^{\min} > 0$, maximum rates $\{x_i^{\max}\}_{i \in \mathcal{N}}$ with groups $\mathcal{N}_1, \dots, \mathcal{N}_K$, tolerances $\varepsilon, \delta > 0$}
\KwOut{allocation $\boldsymbol{x}^\star$}
\tcp{Bisect}
$\lambda^{\mathrm{th}}_i \leftarrow f_i(x^{\min})/x^{\min}$ for all $i \in \mathcal{N}$\;
$\lambda^- \leftarrow 0$;\quad $\lambda^+ \leftarrow \max_{i \in \mathcal{N}} \lambda^{\mathrm{th}}_i$\;
\While{$\lambda^+ - \lambda^- > \varepsilon$}{
    $\lambda \leftarrow (\lambda^- + \lambda^+)/2$\;
    \ForEach{$i \in \mathcal{N}$}{
        \leIf{$\lambda < \lambda^{\mathrm{th}}_i$}{$x_i \leftarrow \bigl[(f_i')^{-1}(\lambda)\bigr]_{x^{\min}}^{x_i^{\max}}$}{$x_i \leftarrow 0$}
    }
    $S \leftarrow \sum_{i \in \mathcal{N}} x_i$\;
    \lIf{$0 \le B - S \le \delta$}{\Return{$\boldsymbol{x}$} \tcp*[f]{fast-path}}
    \leIf{$S > B$}{$\lambda^- \leftarrow \lambda$}{$\lambda^+ \leftarrow \lambda$}
}
\tcp{Index}
sort each group $\mathcal{N}_k$ in the priority order\;
\tcp{Choose}
\tcp{$K = 2$; for $K = 1$ and $K > 2$, see Section~\ref{sec:choosing}}
$n^{\max} \leftarrow \lfloor B / x^{\min} \rfloor$\;
$\boldsymbol{c}^\star \leftarrow \textsc{Choose}(\mathcal{N}_1, \mathcal{N}_2, n^{\max})$ \tcp*{Algorithm~\ref{alg:choose}}
\Return{$x^\star_i \leftarrow$ solution of $(P_{\mathcal{I}^\star_{\boldsymbol{c}^\star}})$ for $i \in \mathcal{I}^\star_{\boldsymbol{c}^\star}$, and $x^\star_i \leftarrow 0$ otherwise}\;
\caption{Bisect-Index-Choose (BIC).}
\label{alg:bic}
\end{algorithm}

The inner loop (lines~\ref{ln:ch-while} to~\ref{ln:ch-dec}) scans $c_2$ downwards, from a count that is at least $\hat c_2(c_1)$. For $c_1 = 0$, it starts from $n_2$ (line~\ref{ln:ch-init}). For larger $c_1$, it starts from $\hat c_2(c_1 - 1)$, which is at least $\hat c_2(c_1)$ by Theorem~\ref{thm:dr}(b). In both cases, the start is capped by the maximum number of EVs that the remaining budget can switch on, $n^{\max} - c_1$ (line~\ref{ln:ch-cap}). Due to Theorem~\ref{thm:dr}(a), when $V_{(c_1, c_2 - 1)} < V_{(c_1, c_2)}$, the best count of group~2 cannot be less than the current $c_2$, so the scan stops at this condition (line~\ref{ln:ch-test}), and since it started at or above $\hat c_2(c_1)$, the current $c_2$ is $\hat c_2(c_1)$. The value $V_{(c_1, \hat c_2(c_1))}$ is recorded, and the outer loop moves to $c_1 + 1$. Since this is done for every feasible $c_1$, the argument of the best recorded value gives $\boldsymbol{c}^\star$ (lines~\ref{ln:ch-if} and~\ref{ln:ch-keep}).

Each value $V_{(c_1, c_2)}$ is computed by solving its restriction when it is first needed. Each $c_1$ adds at most two new values at the test of line~\ref{ln:ch-test}, and $c_2$ decreases at most $n_2$ times in total, so \textsc{Choose} solves at most $2(n_1 + 1) + n_2$ restrictions.

For $K = 1$, Theorem~\ref{thm:dr}(a) alone suffices. With $c$ a single count, it states that $V_{c+1} > V_c$ holds exactly when $c < c^\star = \hat c_1$, which for $K = 1$ is a single number. This test is true for every count below $c^\star$ and false from $c^\star$ on. A bisection on the count over $\{0, \dots, \min(n, n^{\max})\}$ therefore finds $c^\star$ with at most $2\lceil \log_2(n+1) \rceil$ restrictions.

The generalization to any $K > 2$ is straightforward. Since Theorem~\ref{thm:dr} holds for any $K$, groups $K-1$ and $K$ play the roles of groups 1 and 2 in Algorithm~\ref{alg:choose} once $c_1, \dots, c_{K-2}$ are fixed. \textsc{Choose} therefore runs once for every feasible choice of $c_1, \dots, c_{K-2}$, with $n^{\max}$ reduced by $c_1 + \dots + c_{K-2}$, and the best of the returned cardinality vectors is $\boldsymbol{c}^\star$.

\subsection{The Algorithm, Correctness, and Runtime}\label{sec:algorithm}

Algorithm~\ref{alg:bic} gives the full BIC procedure for $K = 2$; for other $K$, its call to \textsc{Choose} is replaced by the choosing phase of Section~\ref{sec:choosing}. Each bisection stops when the allocation uses between $B - \delta$ and $B$, or when the bracket on $\lambda$ is narrower than $\varepsilon$, where $\delta$ and $\varepsilon$ are several orders of magnitude smaller than $B$ and $L$, respectively.

\begin{theorem}[Correctness of BIC]\label{thm:correct}
With exact bisections, BIC returns a global maximizer of \eqref{eq:P}. With the tolerance $\delta$, an allocation returned at the fast-path, at a multiplier $\lambda$, is at most $\lambda\delta$ below the optimal value of \eqref{eq:P}.
\end{theorem}

\begin{proof}
BIC returns at the fast-path or after the choosing phase.

At the fast-path, the allocation $\boldsymbol{x}(\lambda)$ satisfies $B - \delta \le S(\lambda) \le B$. By Theorem~\ref{thm:bisection}, applied with the budget $S(\lambda)$, it is a global maximizer of \eqref{eq:P} for a budget within $\delta$ of $B$. Since $\boldsymbol{x}(\lambda)$ maximizes $\sum_{i \in \mathcal{N}} (f_i(x_i) - \lambda x_i)$ over the admissible rates, its objective is at most $\lambda\delta$ below the optimal value of \eqref{eq:P}.

After the choosing phase, the optimal value of \eqref{eq:P} is $V^\star = \max_{\boldsymbol{c} \in \mathcal{C}} V_{\boldsymbol{c}}$ by~\eqref{eq:decomposition} and Theorem~\ref{thm:prefix}. By Theorem~\ref{thm:dr}, the choosing phase returns a $\boldsymbol{c}^\star$ that attains this maximum (Section~\ref{sec:choosing}). The solution of $(P_{\mathcal{I}^\star_{\boldsymbol{c}^\star}})$, with zero rate for every EV outside $\mathcal{I}^\star_{\boldsymbol{c}^\star}$, therefore attains $V^\star$.
\end{proof}

\begin{remark}[Per-period and daily optimality]\label{rem:daily}
Theorem~\ref{thm:correct} concerns the optimality in a single control period. In the online operation of Section~\ref{sec:formulation}, an optimal allocation in every period need not give the best schedule for the day, since future arrivals are unknown. Sections~\ref{sec:cap} and~\ref{sec:dr} evaluate how the per-period optimality of BIC translates into the energy delivered over a day.
\end{remark}

\begin{proposition}[Runtime of BIC]\label{prop:complexity}
BIC runs in $O(n \log n \log \frac{L}{\varepsilon})$ time for $K = 1$ and in $O(n^K \log \frac{L}{\varepsilon})$ time for $K \ge 2$.
\end{proposition}

\begin{proof}
We assume that each evaluation of $f_i$, $f_i'$, and $(f_i')^{-1}$, and each comparison in the priority order, takes $O(1)$ time. We count the cost of each phase.
\begin{itemize}
\item \textit{Bisection:} computing the thresholds takes $O(n)$, and the bisection on $[0, L]$ takes at most $\lceil \log_2(L/\varepsilon) \rceil$ iterations of cost $O(n)$, i.e., $O(n \log(L/\varepsilon))$ in total.
\item \textit{Indexing:} sorting the groups takes $O(n \log n)$.
\item \textit{Choosing:} at $\lambda = L > f_i'(x^{\min})$ (Appendix~\ref{app:closed}), every EV of a restriction is at $x^{\min}$, and the total $|\mathcal{I}|\,x^{\min}$ is at most $B$. The multiplier of each restriction therefore lies in $[0, L]$, and its bisection costs $O(n \log(L/\varepsilon))$. By Section~\ref{sec:choosing}, the phase solves $O(\log n)$ restrictions for $K = 1$ and $O(n^{K-1})$ for $K \ge 2$, namely $O(n^{K-2})$ runs of \textsc{Choose} with $O(n)$ restrictions each, which gives $O(n \log n \log(L/\varepsilon))$ and $O(n^K \log(L/\varepsilon))$, respectively.
\end{itemize}
The choosing phase dominates the other two phases, so its cost is the runtime of BIC.
\end{proof}

BIC therefore runs in polynomial time for a fixed $K$, which is small in practice (Section~\ref{sec:formulation}). When the budget is not in an Everett gap, the fast-path stops BIC after the bisection phase, at a cost of $O(n \log(L/\varepsilon))$ for any $K$.

\section{Experimental Setup}\label{sec:setup}

We evaluate BIC numerically in four respects: the optimality of its allocations, its computation time and scalability, its execution on a low-cost microcontroller with limited memory and computing power, and its performance in the online operation of a charging station on a real-world dataset of charging sessions.

\subsection{Utility Functions and Charging Rates}\label{sec:setup-utility}

Throughout, we instantiate the utility functions from the laxity-based family of sLLF~\cite{chen2021smoothed}, $f_i(x) = x_i^{\max}\log(\ell_i - 1 + x/x_i^{\max})$ up to the constant that enforces $f_i(0) = 0$, which is the first row of Table~\ref{tab:families}. Here $\ell_i$ is the laxity of EV $i$ at the current control period, measured in control periods as the difference between its remaining parking time and the minimum time needed to deliver its remaining requested energy at its maximum rate. Laxity is recomputed every control period and floored so that $\ell_i - 1 \ge 10^{-6}$, which keeps the logarithm well defined. Within each group, these utilities satisfy Assumption~\ref{as:priority} with sorting key $\ell_i$. We take a control period length of 1 minute and $x^{\min} = 4.16$~kW, the 6~A minimum current of IEC~61851~\cite{iec61851} and SAE~J1772~\cite{saej1772} at a 0.4~kV three-phase supply. Each EV is rated $x_i^{\max} = 11$ or 22~kW with equal probability, which gives two groups.

\subsection{Benchmarks and Implementation}\label{sec:setup-methods}

We compare BIC with the exact solution of \eqref{eq:P}, computed with the commercial MIP solver MOSEK~\cite{mosek} via CVXPY~\cite{diamond2016cvxpy}, and with two heuristic benchmarks (HBs). HB1 is adapted from the rounding rule of~\cite{kisacikoglu2018distributed}. It solves~\eqref{eq:P_bar}, switches off every EV whose relaxed rate falls below $x^{\min}$, and solves the restriction~\eqref{eq:P_J} on the remaining EVs. Unlike the original rule, HB1 never rounds a rate up to $x^{\min}$, since that could exceed the budget. HB2 is adapted from the fallback rule of~\cite[fn.~3]{lee2021adaptive}. It switches on the $\min(n, n^{\max})$ EVs of lowest laxity and solves the restriction on them. Both heuristics use the objective $\sum_{i} f_i$. BIC uses the parameters $\delta = 10^{-6}B$ and $\varepsilon = 10^{-13}$, and the solver runs with its default settings, with a time limit of 10~s per instance in Section~\ref{sec:opt}, after which its best allocation is kept.

All methods are implemented in Python and run on a single thread of an Intel Core i7-10750H CPU at 2.60~GHz. For the microcontroller experiments, we implement BIC in C with 64-bit floating-point arithmetic. It runs on one core of an Arm Cortex-M33 microcontroller (RP2350~\cite{rp2350datasheet}, 150~MHz, 520~kB of on-chip memory, Wi-Fi), without an operating system or solver. The C implementation makes the same decisions as Algorithm~\ref{alg:bic} and, to shorten the computation time, reuses computations across bisection steps and restrictions, which the closed form of the sLLF utilities allows.

\subsection{Synthetic Instances and Real-World EV Data}\label{sec:setup-data}

We consider two settings. In the first, we study a single control period on synthetic instances, whose laxities and budget we choose to probe specific aspects of BIC: its optimality against the MIP solver and the heuristics (Section~\ref{sec:opt}), its scaling with the number of EVs (Section~\ref{sec:scal}), and its execution on the microcontroller (Section~\ref{sec:pico}).

In the second, we simulate the online operation of a charging station on the public dataset of~\cite{debont2024evcharging}, which records the arrival and departure times and the delivered energy of the charging sessions at a large office parking lot in Utrecht. We take the delivered energy as the energy request, draw the maximum rates as described in Section~\ref{sec:setup-utility}, and cap each request at what the EV can receive during its stay to make every request feasible. We simulate separately each of the 714 days between August 2021 and October 2024 that have at least ten sessions. With this dataset, Section~\ref{sec:cap} holds the budget constant, and Section~\ref{sec:dr} adds a demand response event.

\section{Results and Discussion}\label{sec:numerics}

\subsection{Optimality on Synthetic Instances}\label{sec:opt}

We first evaluate solution quality on synthetic instances with $n = 100$ EVs, built in two families, each to expose the weakness of one heuristic. Both families contain a block of EVs with identical utilities, which share one maximum rate and create a single Everett gap, in which we place the budget. In the HB1-adversarial family, the relaxed solution spreads the power of this gap over the whole block at rates just below $x^{\min}$, so HB1 switches the block off. In the HB2-adversarial family, a few urgent EVs precede the block in the priority order, and many EVs with large laxity follow it. HB2 switches on as many EVs as the budget allows, so it gives $x^{\min}$ to these low-priority EVs and takes power away from the urgent ones.

We vary the share of identical EVs from 0 to $100\%$ in steps of $20\%$, where the share of $0\%$ contains no Everett gap and serves as a control. We also vary the budget ratio $\sigma = B/(n \cdot 11~\text{kW})$, the budget relative to all EVs charging at 11~kW, from 0.1 to 0.8. Each setting has up to ten seeded instances, 815 in total over both families. Some combinations of share and $\sigma$ cannot be constructed, since the identical EVs cannot then create an Everett gap that contains the budget, for example when all EVs are identical and the budget exceeds the power they draw together at $x^{\min}$. The full construction of both families is given in the code (footnote~\ref{fn:code}).

\begin{figure}[t]
\centering
\includegraphics[width=\columnwidth]{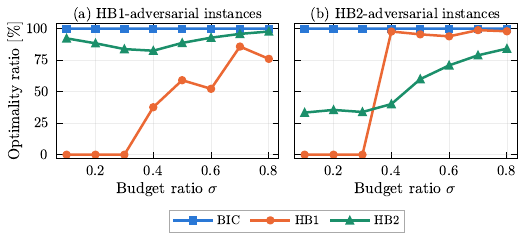}
\caption{Optimality ratio, the total utility of an allocation divided by that of the MIP solution, on (a) HB1-adversarial and (b) HB2-adversarial instances with $n=100$ EVs. Each point reports the worst setting at that $\sigma$, i.e., the lowest median ratio over the shares of identical EVs. Ratios above $100\%$, at most $100.9\%$, arise when the MIP solver reaches its time limit before the optimum, and are shown as $100\%$.}
\label{fig:optimality}
\vspace{2mm}
\includegraphics[width=\columnwidth]{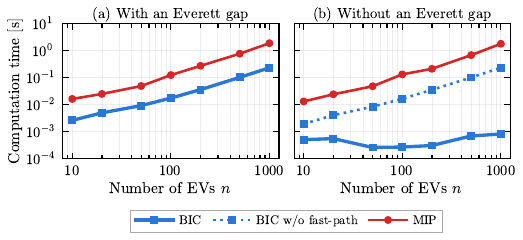}
\caption{Computation time of BIC and of the MIP solver on a single thread, for $n=10$ to $1000$ EVs at $\sigma = 0.38$, (a) with and (b) without an Everett gap; in (b), BIC without the fast-path runs all three phases. Points report the median over five random seeds.}
\label{fig:scalability}
\vspace{-0.8mm}
\end{figure}

Fig.~\ref{fig:optimality} reports, for each method and $\sigma$, the worst setting, i.e., the lowest median optimality ratio over the shares of identical EVs. BIC attains global optimality on every instance, as Theorem~\ref{thm:correct} guarantees. Each heuristic, in contrast, fails badly on the family built against it. HB1 loses the entire utility at $\sigma \le 0.3$ in both families, since all relaxed rates then fall below $x^{\min}$ and HB1 switches off every EV. HB2 falls to $33\%$ on its adversarial family, and this worst case occurs at a share of $0\%$, without a gap, so its loss comes from switching on too many EVs. These results show that simple fleets of urgent and non-urgent EVs, or of EVs with similar urgency, suffice to make either heuristic lose most of the utility, whereas BIC remains optimal on all of them.

\subsection{Computation Time on Synthetic Instances}\label{sec:scal}

To measure the computation time of BIC, we generate instances with $n = 10$ to $1000$ EVs at $\sigma = 0.38$. For each $n$, we create two instances that are identical except for the laxity of a single EV. In the first, the budget lies inside an Everett gap, so the indexing and choosing phases run, whereas in the second, it is not in an Everett gap, so BIC terminates through the fast-path. At this budget ratio, $n^{\max} \ge n$, so the cap $n^{\max}$ never shortens the search of the choosing phase. Fig.~\ref{fig:scalability} reports median times over five random seeds. With a gap, BIC solves 100 EVs in 18~ms and 1000 EVs in 0.23~s, 7.1 and 8.1 times faster than the MIP solver, whose time includes the modeling in CVXPY. Without a gap, BIC needs at most 0.9~ms at every size, up to 2200 times faster than the solver. Here, the fast-path pays off: without it, BIC would run all three phases and take as long as with a gap, so the fast-path makes BIC 62 times faster at 100 EVs and 280 times faster at 1000 EVs. In every instance, the objective of BIC is at least that of the solver, up to a relative $10^{-9}$.

\begin{figure}[t]
\centering
\includegraphics[width=\columnwidth]{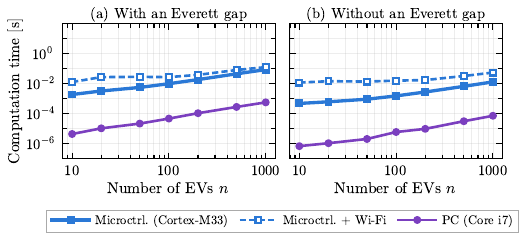}
\caption{Median computation time of BIC on an Arm Cortex-M33 microcontroller (RP2350, 150~MHz) for the instances of Fig.~\ref{fig:scalability}, (a) with and (b) without an Everett gap. ``Microctrl.\ + Wi-Fi'' is the round trip, including the communication with the PC over Wi-Fi, and ``PC (Core i7)'' runs the same C code on the PC as a reference. Points report the median over the five random seeds of Fig.~\ref{fig:scalability}.}
\label{fig:pico}
\vspace{2mm}
\includegraphics[width=\columnwidth]{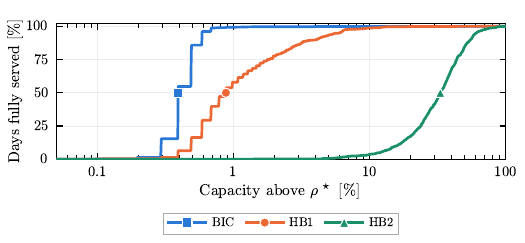}
\caption{Constant power budget over 714 days of charging sessions: capacity that each method needs above the offline minimum $\rho^\star$ (Section~\ref{sec:cap}) to deliver every request. For each method, the curve shows the share of days on which the method delivers every request with at most the given relative capacity above $\rho^\star$, and the marker shows the median.}
\label{fig:cap-capacity}
\vspace{-0.8mm}
\end{figure}

\begin{figure*}[t]
\begin{minipage}[t]{\columnwidth}
\centering
\includegraphics[width=\columnwidth]{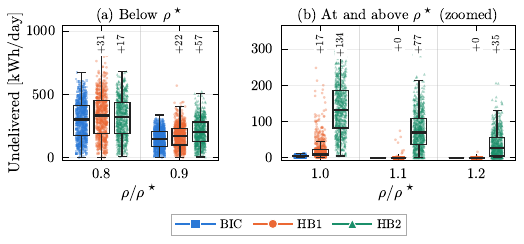}
\caption{Constant power budget over 714 days of charging sessions: energy left undelivered per day at a capacity ratio of 0.8 to 1.2 times each day's $\rho^\star$, (a) below and (b) at and above $\rho^\star$. Each dot is one day, boxes show the quartiles, and the labels give the mean undelivered energy of HB1 and HB2 in excess of BIC, in kWh.}
\label{fig:cap-unserved}
\end{minipage}\hfill
\begin{minipage}[t]{\columnwidth}
\centering
\includegraphics[width=\columnwidth]{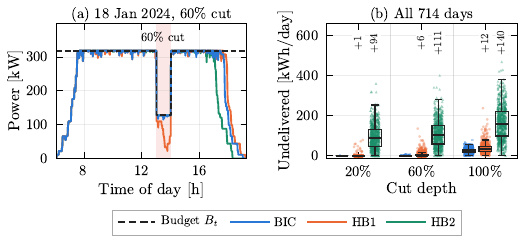}
\caption{Demand response with a one-hour cut of the normal budget $1.1\,\rho^\star B^{\max}$ (Section~\ref{sec:cap}). (a) Budget and aggregate charging power on 18 January 2024, the busiest day of the dataset (159 sessions), with a $60\%$ cut from 13:00 to 14:00. (b) Energy left undelivered per day over 714 days, for cuts of $20\%$, $60\%$, and $100\%$. Dots, boxes, and labels as in Fig.~\ref{fig:cap-unserved}.}
\label{fig:dr-combined}
\end{minipage}
\vspace{-1.7mm}
\end{figure*}

\subsection{Implementation on a Microcontroller}\label{sec:pico}

We now test whether BIC can run on the resource-constrained microcontroller described in Section~\ref{sec:setup-methods}. In this test, the microcontroller acts as the station controller, and the PC emulates the rest of the charging station, which holds the data of the connected EVs. The two communicate over Wi-Fi. We reuse the instances of Section~\ref{sec:scal}. For each instance, the PC sends the budget and the remaining parking time, remaining energy demand, and maximum rate of every EV, and the microcontroller computes the utilities, runs BIC, and returns the rates, as a station controller would in every control period. We record the computation time with the on-chip timer of the microcontroller and the round-trip time on the PC, and run the same C code on the PC as a reference.

Fig.~\ref{fig:pico} reports the median times. All allocations computed on the microcontroller coincide with those of the Python implementation, so the limited hardware costs no accuracy. Even in the worst case, with a gap and 1000 EVs, the microcontroller needs less than 90~ms, about $0.15\%$ of a one-minute control period, and the round trip including Wi-Fi takes 0.13~s. Without a gap, the fast-path keeps the computation below 15~ms at every size. The microcontroller is 150 to 700 times slower than the PC running the same code, but its time grows with the number of EVs at almost the same rate, so the slowdown comes from the hardware, not from the algorithm. Overall, BIC runs on a low-cost edge device with a wide margin: it uses a small fraction of a control period, which leaves ample time for communication and other tasks.

\subsection{Real-World Case Study: Constant Power Budget}\label{sec:cap}

In the first case study on the real-world dataset, we hold the budget constant over each day. We express each day's budget by the capacity ratio $\rho = B/B^{\max}$, where $B^{\max} = 22~\text{kW} \cdot n^{\mathrm{peak}}$ and $n^{\mathrm{peak}}$ is the largest number of EVs connected at the same time on that day. As a reference, $\rho^\star$ is the smallest constant $\rho$ at which an \emph{oracle}, i.e., a planner with full knowledge of the day, could deliver every request. It is computed by linear programming without the minimum rate and is thus a lower bound on the capacity that any scheduler needs.

We run two experiments. First, for each method and day, we find by bisection over $\rho$ the smallest capacity ratio at which the method delivers all requested energy, to a resolution of $0.1\%$ of $\rho^\star$. Second, we fix each day's capacity ratio at 0.8 to 1.2 times its $\rho^\star$ and record the energy left undelivered.

Fig.~\ref{fig:cap-capacity} shows the results of the first experiment. For each method, the curve gives the share of days on which the method delivers every request with at most a given capacity above $\rho^\star$, so a curve further to the left means less capacity. BIC needs only slightly more capacity than the oracle, a median of $0.4\%$ above $\rho^\star$ and at most $7.1\%$ on any day, whereas HB1 needs up to $21.1\%$ and HB2 up to $99.3\%$. On all but one day, BIC needs no more capacity than either heuristic (see Remark~\ref{rem:daily}).

Fig.~\ref{fig:cap-unserved} shows the results of the second experiment. Each dot is the energy left undelivered on one day, and the labels give the mean excess of HB1 and HB2 over BIC. From $1.1\,\rho^\star$ on, BIC delivers every request on every day, whereas both heuristics still leave energy undelivered, HB2 by a wide margin. At $\rho^\star$, BIC leaves about a quarter of the energy that HB1 leaves undelivered and a small fraction of that of HB2, and below $\rho^\star$, where no method can deliver every request, BIC still leaves the least.

In summary, although BIC decides online, it stays close to the oracle: it needs at most $1\%$ more capacity than $\rho^\star$ on $99.4\%$ of the days, against $58.0\%$ for HB1 and none for HB2. Under a limited budget, it also leaves the least energy undelivered, at $\rho^\star$ about a quarter of that of HB1 and less than $4\%$ of that of HB2.

\subsection{Real-World Case Study: Demand Response}\label{sec:dr}

We then model a demand response call, in which a grid operator or aggregator asks the station to reduce its power for a limited time, for example to relieve a congested network or to balance supply and demand~\cite{wang2025online}. The station must then charge the connected EVs with less power during the call, and can use the full budget again afterwards. We simulate this as follows: on each day, the normal budget $B_0 = 1.1\,\rho^\star B^{\max}$ is cut by a fraction of $B_0$ for one hour from 13:00, and restored at 14:00. We start from $1.1\,\rho^\star$ because, from this budget on, BIC delivers every request on every day (Section~\ref{sec:cap}), so any energy that BIC leaves undelivered is caused by the cut. We consider cuts of $20\%$, $60\%$, and $100\%$ of $B_0$.

Fig.~\ref{fig:dr-combined}(a) shows the busiest day, 18 January 2024, with a $60\%$ cut. BIC keeps the aggregate power at or close to the budget through the cut. HB1 draws well below it, because it switches off every EV whose relaxed rate falls below $x^{\min}$ and the remaining EVs cannot absorb the freed power. HB2 keeps the power at the budget but spreads it over as many EVs as possible, and leaves 164.7~kWh undelivered. BIC delivers every request, and so does HB1, which catches up after the cut because the EVs it switched off are still connected.

Fig.~\ref{fig:dr-combined}(b) summarizes the results for all 714 days. With BIC, the station absorbs cuts of up to $60\%$ with virtually no effect on its users, and even when the budget is cut to zero for the whole hour, it still delivers $98.5\%$ of the requested energy. The heuristics fall behind exactly when power is scarce. The gap between HB1 and BIC widens as the cut deepens, because more relaxed rates fall below $x^{\min}$ and HB1 switches these EVs off. HB2 loses energy even without a cut, because it spreads the power over too many EVs, and the cut only adds to this loss. In conclusion, the minimum rate matters most when the budget is tight, and by handling it optimally, BIC lets a station offer substantial flexibility to the grid with little effect on the charging of its EVs.

\section{Conclusion}\label{sec:conclusion}

This paper has presented BIC, a solver-free algorithm that, under mild assumptions on the utilities, computes a globally optimal allocation of a station's power budget among EVs with different maximum rates under the minimum charging rate of the charging standards. For $n$ EVs, BIC runs in $O(n \log n)$ time when all EVs share one maximum rate and in $O(n^2)$ time for two, times a logarithmic factor in the numerical precision. On synthetic instances, BIC attains global optimality in every case, whereas two heuristics adapted from the literature fall to $0\%$ and $33\%$ of the optimal utility in their worst cases. With 1000 EVs, BIC is up to three orders of magnitude faster than a commercial MIP solver, and on a low-cost microcontroller its runtime is less than 90~ms. On 714 days of real charging sessions, BIC needs no more capacity than either heuristic on all but one day, and stays close to the oracle.

BIC has a few limitations that future work can address. It optimizes each control period separately, so its schedule over a day need not be optimal (Remark~\ref{rem:daily}). Its runtime grows as $n^K$ with the number $K$ of distinct maximum rates, although $K$ is small in practice. It requires the utilities to satisfy Assumptions~\ref{as:regularity} and~\ref{as:priority}. It assumes that the supply and all EVs have the same phase configuration, either all single-phase or all three-phase, so that one minimum rate and one budget apply. Finally, its only constraints are the admissible rates and this budget, and further constraints, such as per-phase or network limits, cannot be added directly.

\appendix
\subsection{Utility Families with a Key}\label{app:assumptions}

Let the utilities of each group $\mathcal{N}_k$ come from a common family, $f_i(x) = h_k(x; \kappa_i)$ for $i \in \mathcal{N}_k$, where the scalar $\kappa_i$ is the key of EV $i$ and $h_k(0; \kappa) = 0$. The family may differ between groups, for example through $\bar{x}_k$. Both assumptions hold under two conditions. First, $h_k(\cdot\,; \kappa)$ satisfies Assumption~\ref{as:regularity} on $[0, \bar{x}_k]$ for every key, which gives Assumption~\ref{as:regularity} directly. Second, the marginal utility $\partial h_k / \partial x\,(x; \kappa)$ is monotone in $\kappa$, in the same direction for every $x \in [0, \bar{x}_k]$. Then Assumption~\ref{as:priority} holds with the EVs sorted by their keys: by descending $\kappa_i$ if the marginal utility increases in $\kappa$, and by ascending $\kappa_i$ if it decreases. Indeed, for $i \le j$ in this order, $f_i' \ge f_j'$ on $[0, \bar{x}_k]$. So $f_i - f_j$ is $0$ at $x = 0$ and non-decreasing, hence $f_i \ge f_j$.

\subsection{Derivation of the Closed Form~\eqref{eq:chi}}\label{app:closed}

Fix $i \in \mathcal{N}$ and $\lambda \ge 0$, and recall from Section~\ref{sec:bisection} that \eqref{eq:chi_argmax} compares the point $0$, with value $0$, with the projection onto $[x^{\min}, x_i^{\max}]$. At $x^{\min}$, the objective of \eqref{eq:chi_argmax} takes the value
\begin{equation*}
f_i(x^{\min}) - \lambda x^{\min} = x^{\min}(\lambda^{\mathrm{th}}_i - \lambda).
\end{equation*}
\begin{itemize}
\item If $\lambda < \lambda^{\mathrm{th}}_i$, this value is positive. The projection is the best point of the interval, so its value is at least as large, and it beats the point $0$.
\item If $\lambda \ge \lambda^{\mathrm{th}}_i$: strict concavity and $f_i(0) = 0$ give $f_i(x^{\min}) > x^{\min} f_i'(x^{\min})$, i.e., $\lambda^{\mathrm{th}}_i > f_i'(x^{\min})$, so $\lambda > f_i'(x^{\min})$. Since $f_i'$ is decreasing, the objective decreases on the whole interval, so the projection equals $x^{\min}$ and its value $x^{\min}(\lambda^{\mathrm{th}}_i - \lambda)$ is non-positive. The point $0$ is therefore optimal, and at $\lambda = \lambda^{\mathrm{th}}_i$, where both candidates attain the value $0$, \eqref{eq:chi} selects the point $0$.
\end{itemize}
Hence \eqref{eq:chi} gives a maximizer in \eqref{eq:chi_argmax}.

\subsection{Proof of Theorem~\ref{thm:prefix}}\label{app:thm2}

\begin{proof}
We transform any on-set $\mathcal{I}$ with cardinality vector $\boldsymbol{c}$ into $\mathcal{I}^\star_{\boldsymbol{c}}$ by a sequence of swaps within groups and show that no swap decreases the value.

Let $\mathcal{I} \ne \mathcal{I}^\star_{\boldsymbol{c}}$ have cardinality vector $\boldsymbol{c}$. Since both on-sets switch on $c_k$ EVs of each group $\mathcal{N}_k$, there exist a group $\mathcal{N}_k$, an EV $i \in (\mathcal{I}^\star_{\boldsymbol{c}} \setminus \mathcal{I}) \cap \mathcal{N}_k$, and an EV $j \in (\mathcal{I} \setminus \mathcal{I}^\star_{\boldsymbol{c}}) \cap \mathcal{N}_k$. Since $i$ is among the first $c_k$ EVs of $\mathcal{N}_k$ in the priority order and $j$ is not, $i < j$. We swap $j$ for $i$, which gives the on-set $\mathcal{I}' := (\mathcal{I} \setminus \{j\}) \cup \{i\}$ with cardinality vector $\boldsymbol{c}$. We show $V(\mathcal{I}') \ge V(\mathcal{I})$.

Let $\boldsymbol{x}^\star$ be an optimal solution to \eqref{eq:P_J} and let $y := x^\star_j \in [x^{\min}, \bar{x}_k]$. On $\mathcal{I}'$, let EV $i$ take the rate of EV $j$ and keep all other rates:
\begin{equation*}
\hat{x}_i := y, \qquad \hat{x}_{r} := x^\star_{r} \ \ \forall r \in \mathcal{I} \setminus \{j\}.
\end{equation*}
This allocation is feasible for $(P_{\mathcal{I}'})$: $y \in [x^{\min}, \bar{x}_k] = [x^{\min}, x_i^{\max}]$, since $i$ and $j$ belong to the same group, and the total power is unchanged, so the budget still holds. In the objective, all terms cancel except those of EVs $i$ and $j$:
\begin{equation*}
\sum_{r \in \mathcal{I}'} f_{r}(\hat{x}_{r}) - \sum_{r \in \mathcal{I}} f_{r}(x^\star_{r}) = f_i(y) - f_j(y) \ \ge\ 0,
\end{equation*}
where the inequality follows from the priority order~\eqref{eq:order}, since $i < j$ belong to $\mathcal{N}_k$ and $y \in [x^{\min}, \bar{x}_k]$. Hence $V(\mathcal{I}') \ge \sum_{r \in \mathcal{I}'} f_{r}(\hat{x}_{r}) \ge V(\mathcal{I})$.

Each swap replaces an EV outside $\mathcal{I}^\star_{\boldsymbol{c}}$ by one inside it, so $|\mathcal{I} \setminus \mathcal{I}^\star_{\boldsymbol{c}}|$ decreases by one. After at most $\sum_k c_k$ swaps, $\mathcal{I}$ becomes $\mathcal{I}^\star_{\boldsymbol{c}}$. As no swap lowers the value, $V(\mathcal{I}^\star_{\boldsymbol{c}}) \ge V(\mathcal{I})$.
\end{proof}

\subsection{Proof of Theorem~\ref{thm:dr}}\label{app:dr}

\begin{lemma}\label{lem:submod}
Let $\mathcal{I}_1, \mathcal{I}_2 \subseteq \mathcal{N}$ with $|\mathcal{I}_1 \cup \mathcal{I}_2| \le n^{\max}$. Then
\begin{equation*}
V(\mathcal{I}_1 \cup \mathcal{I}_2) + V(\mathcal{I}_1 \cap \mathcal{I}_2) \le V(\mathcal{I}_1) + V(\mathcal{I}_2).
\end{equation*}
\end{lemma}

\begin{proof}
Since \eqref{eq:P_J} is convex with linear constraints, strong duality holds, and dualizing its budget constraint gives
\begin{equation*}
V(\mathcal{I}) = \min_{\lambda \ge 0} \mathcal{L}_{\mathcal{I}}(\lambda), \qquad \mathcal{L}_{\mathcal{I}}(\lambda) := \lambda B + \sum_{i \in \mathcal{I}} \varphi_i(\lambda),
\end{equation*}
with $\varphi_i(\lambda) := \max_{x \in [x^{\min}, x_i^{\max}]} \{ f_i(x) - \lambda x \}$. Let $\lambda_1$ and $\lambda_2$ attain the minima for $\mathcal{I}_1$ and $\mathcal{I}_2$, with $\lambda_1 \ge \lambda_2$ without loss of generality. Then
\begin{align*}
&V(\mathcal{I}_1 \cup \mathcal{I}_2) + V(\mathcal{I}_1 \cap \mathcal{I}_2) \\
&\quad\le \mathcal{L}_{\mathcal{I}_1 \cup \mathcal{I}_2}(\lambda_1) + \mathcal{L}_{\mathcal{I}_1 \cap \mathcal{I}_2}(\lambda_2) \\
&\quad= \mathcal{L}_{\mathcal{I}_1}(\lambda_1) + \sum_{i \in \mathcal{I}_2 \setminus \mathcal{I}_1} \varphi_i(\lambda_1) + \mathcal{L}_{\mathcal{I}_1 \cap \mathcal{I}_2}(\lambda_2) \\
&\quad\le \mathcal{L}_{\mathcal{I}_1}(\lambda_1) + \sum_{i \in \mathcal{I}_2 \setminus \mathcal{I}_1} \varphi_i(\lambda_2) + \mathcal{L}_{\mathcal{I}_1 \cap \mathcal{I}_2}(\lambda_2) \\
&\quad= V(\mathcal{I}_1) + V(\mathcal{I}_2),
\end{align*}
where the second inequality holds as each $\varphi_i$ is non-increasing, being a maximum of functions that decrease in $\lambda$.
\end{proof}

\begin{lemma}\label{lem:dr}
Let $\boldsymbol{c} \in \mathcal{C}$ with $c_k \ge 1$, and let $l$ be a group with $\boldsymbol{c} + \boldsymbol{e}_l \in \mathcal{C}$, possibly $l = k$. Then
\begin{equation}\label{eq:dr}
V_{\boldsymbol{c} + \boldsymbol{e}_l} - V_{\boldsymbol{c} + \boldsymbol{e}_l - \boldsymbol{e}_k} \le V_{\boldsymbol{c}} - V_{\boldsymbol{c} - \boldsymbol{e}_k}.
\end{equation}
\end{lemma}

\begin{proof}
Recall that $V_{\boldsymbol{c}} = V(\mathcal{I}^\star_{\boldsymbol{c}})$ by Theorem~\ref{thm:prefix}. In both cases below, $\mathcal{I}_1 \cup \mathcal{I}_2 = \mathcal{I}^\star_{\boldsymbol{c} + \boldsymbol{e}_l}$, so Lemma~\ref{lem:submod} applies because $\boldsymbol{c} + \boldsymbol{e}_l \in \mathcal{C}$.
\begin{itemize}
\item $l \ne k$: Let $\mathcal{I}_1 = \mathcal{I}^\star_{\boldsymbol{c} - \boldsymbol{e}_k + \boldsymbol{e}_l}$ and $\mathcal{I}_2 = \mathcal{I}^\star_{\boldsymbol{c}}$. Then $\mathcal{I}_1 \cup \mathcal{I}_2 = \mathcal{I}^\star_{\boldsymbol{c} + \boldsymbol{e}_l}$ and $\mathcal{I}_1 \cap \mathcal{I}_2 = \mathcal{I}^\star_{\boldsymbol{c} - \boldsymbol{e}_k}$, and Lemma~\ref{lem:submod} gives~\eqref{eq:dr}.
\item $l = k$: Let $j$ be the $(c_k + 1)$-th EV of $\mathcal{N}_k$, $\mathcal{I}_1 = \mathcal{I}^\star_{\boldsymbol{c}}$ and $\mathcal{I}_2 = \mathcal{I}^\star_{\boldsymbol{c} - \boldsymbol{e}_k} \cup \{j\}$. Then $\mathcal{I}_1 \cup \mathcal{I}_2 = \mathcal{I}^\star_{\boldsymbol{c} + \boldsymbol{e}_k}$ and $\mathcal{I}_1 \cap \mathcal{I}_2 = \mathcal{I}^\star_{\boldsymbol{c} - \boldsymbol{e}_k}$. Since $\mathcal{I}_2$ has the cardinality vector $\boldsymbol{c}$, Theorem~\ref{thm:prefix} gives $V(\mathcal{I}_2) \le V_{\boldsymbol{c}}$, and Lemma~\ref{lem:submod} gives~\eqref{eq:dr}.\qedhere
\end{itemize}
\end{proof}

\begin{proof}[Proof of Theorem~\ref{thm:dr}]
By~\eqref{eq:cardinalities}, $\boldsymbol{c}(k; q) \in \mathcal{C}$ exactly for $q = 0, \dots, \bar{q}$, where $\bar{q} := \min\{n_k, n^{\max} - \sum_{m \ne k} c_m\}$. For $q = 1, \dots, \bar{q}$, let
\begin{equation*}
g(q) := V_{\boldsymbol{c}(k; q)} - V_{\boldsymbol{c}(k; q - 1)}
\end{equation*}
be the gain of switching on the $q$-th EV of $\mathcal{N}_k$, so that $V_{\boldsymbol{c}(k; q)} = V_{\boldsymbol{c}(k; 0)} + g(1) + \dots + g(q)$.

For part~(a), Lemma~\ref{lem:dr} with $l = k$ gives $g(1) \ge g(2) \ge \dots \ge g(\bar{q})$. Let $p$ be the number of positive gains. Then $g(q) > 0$ for $q \le p$ and $g(q) \le 0$ for $q > p$. Hence $V_{\boldsymbol{c}(k; q)}$ increases strictly up to $q = p$ and does not increase afterwards, so its smallest maximizer is $\hat c_k(\boldsymbol{c}) = p$. In particular, $g(c_k) = V_{\boldsymbol{c}} - V_{\boldsymbol{c} - \boldsymbol{e}_k}$ is positive exactly when $c_k \le \hat c_k(\boldsymbol{c})$.

For part~(b), let $\boldsymbol{c}' := \boldsymbol{c} + \boldsymbol{e}_l$, and define $\bar{q}'$ and $g'(q)$ as $\bar{q}$ and $g(q)$ with $\boldsymbol{c}'$ in place of $\boldsymbol{c}$. Since $l \ne k$, $\boldsymbol{c}'(k; q) = \boldsymbol{c}(k; q) + \boldsymbol{e}_l$, so Lemma~\ref{lem:dr} with $l \ne k$, applied at $\boldsymbol{c}(k; q)$, gives $g'(q) \le g(q)$ for $q = 1, \dots, \bar{q}'$. As in part~(a), $\hat c_k(\boldsymbol{c}') = p'$, where $p'$ is the number of positive gains $g'(q)$, and these are $g'(1), \dots, g'(p')$. If $p' \ge 1$, then $g(p') \ge g'(p') > 0$, and since the positive gains of $g$ are $g(1), \dots, g(p)$, it follows that $p' \le p$, which also holds trivially if $p' = 0$. Hence $\hat c_k(\boldsymbol{c}') \le \hat c_k(\boldsymbol{c})$.
\end{proof}

\section*{Acknowledgment}

Claude Opus 5.5 and Fable 5.1 (Anthropic) were used for the code, simulations and figures, as a writing editor, and for autoformalization. The author verified all outputs.

\bibliography{bic_bibliog}

\end{document}